\documentclass[11pt,twoside]{article}
\usepackage{./asp2014}

\aspSuppressVolSlug
\resetcounters

\begin{document}

\title{Imaging Spatially Extended Objects with Interferometers: Mosaicking and the Short Spacing Correction}
\author{Brian~S.~Mason$^1$}
\affil{$^1$National Radio Astronomy Observatory, Charlottesville, VA, USA; \email{bmason@nrao.edu}}

\paperauthor{Brian~S.~Mason}{bmason@nrao.edu}{}{NRAO}{Scientific Services}{Charlottesville}{VA}{22903}{USA}
%\paperauthor{Steven~T.~Myers}{smyers@nrao.edu}{}{NRAO}{Scientific Services}{Socorro}{NM}{87801}{USA}

\begin{abstract}
Interferometry is a powerful technique for making sensitive,
high-fidelity images of the sky, but is limited in its ability to
measure extended or diffuse emission.  Better images of extended
astronomical objects can be obtained by mosaicking together many
pointings of the interferometer array. Even better images can be
obtained by combining these data with data from a single-dish
telescope.  This lecture explains commonly practiced techniques for
obtaining and analyzing these observations, and the theory behind
them.
\end{abstract}

% Focus: large-scale imaging & single dish correction. 
% approach: mathematically sound but also practical. present all the important
%  considerations and ideas with pointers to the literature for more.
%  so, a stand-alone reference for doing it.

\section{Introduction}
\label{sec:intro}

An interferometer can make remarkably accurate measurements of the sky
intensity, but the features that these measurements capture is limited
by the length and orientation of the interferometer's baselines. The
absence of information at spatial frequencies higher than that
provided by the interferometer's {\it longest} baselines results in 
limited angular resolution, making features much smaller than the
scale of $\theta_{min} = \lambda/b_{max}$ impossible to distinguish.  
%This situation can sometimes be addressed by increasing the value of
%$b_{max}$, {\it e.g.}, moving some of the antennas to provide longer
%baselines.
The absence of information on angular scales larger than those
measured by the interferometer's {\it shortest} baselines, $\theta >
\theta_{max} \sim (\lambda/b_{min})$, can in many instances be even
more problematic.  The interferometer can sometimes be reconfigured to
measure larger scales by having more, shorter baselines. However,
since the baselines cannot be shorter than the dish diameter $D$,
$\theta_{max}$ can never be larger than $ (\lambda/D)$.  The missing
large-scale information corresponds to {\it missing flux} for
spatially resolved objects; and can give rise, in effect, to variable
local backgrounds that make it challenging to measure integrated flux
densities, even of relatively compact objects.  Since much
astrophysics is done by comparing spatially integrated features at
different wavelengths--- {\it e.g.}, the integrated line intensity in
two transitions, or the integrated continuum flux density at two
widely separated frequencies--- this can be a significant limitation.

% **I would like to have a real-life ALMA SIM with ACA for UV coverage
%  **and synth beam purposes** --> i can probably use the sim i did
%  **for my talk.

This lecture will address two of the main methods to overcome this
limitation: acquiring and jointly analyzing data from adjacent
interferometer pointings, a technique known as {\it mosaicking}; and
combining data from a single-dish telescope(s) with the interferometer
data.  In \S~\ref{sec:ekersrots}, I will articulate the problem and
present the mathematical and conceptual foundation for both of these
methods, the Ekers-Rots theorem. We will see that the Ekers-Rots
theorem implies that the shortest baseline $b_{min}$ in fact contains
information about scales $\theta > (\lambda/b_{min})$, and gives a
conceptual scheme to retrieve that information from mosaicked
interferometer observations. Several common approaches for analyzing
interferometric mosaics are discussed in \S~\ref{sec:mosaicking}.
Techniques for combining single dish and interferometer data are
discussed in \S~\ref{sec:combining}, and some common ``real-world''
considerations are discussed in \S~\ref{sec:practical}.

A few general comments are in order. The purpose of this lecture is
to explain the current, generally accepted best practices involved in
interferometric imaging of spatially extended astronomical objects,
and the theory behind these practices.   The
reader is presumed to be familiar with the basic concepts of synthesis
imaging, such as the CLEAN algorithm for
deconvolution.  These foundational topics are covered in
\citet{sissII}, and with greater depth and mathematical rigor in
\citet{thompsonMoranSwenson}. The reader may also find the online
lectures from recent Synthesis Imaging Workshops to be of use
\citep[e.g.][]{siss14web,siss16web}.
The field of astronomical
imaging is dynamic, and consequently there are numerous new and
exciting techniques under development that are applicable to the
problems we consider. While pointers into this literature are
provided, there is no attempt to survey these
approaches systematically, which could constitute an article unto itself.
In this article the clean
deconvolution {\it algorithm} will be denoted as CLEAN;
implementations of CLEAN (or other imaging algorithms) in particular
packages will be indicated in typewriter font--- {\tt clean}, {\tt
  tclean}, or {\tt MOSMEM} for instance. Most of the practical
examples in this lecture use the  CASA package.  At the time
these lecture notes were posted the most recent CASA release was $5.6$. In
this and recent releases, the generally recommended implementation of
CLEAN is the task called {\tt tclean}.

\section{Mosaicking Fundamentals}
\label{sec:ekersrots}

\subsection{The Problem}

There are two, related criteria by which a source can be considered
``large'' from the point of view of synthesis imaging:
\begin{enumerate}
\item The source is large compared to the  scale measured by
  the shortest baseline $b_{min}$: $\theta_{src} > \lambda/b_{min}$.
\item The source is large compared to the antenna primary beam:
  $\theta_{src} > \theta_{B}$.
\end{enumerate}
Here $\theta_{B}$ is the Full-Width at Half Maximum (FWHM) of the
antenna primary beam; $\lambda$ is the wavelength of the observation;
and $\theta_{src}$ is the source size.  As mentioned previously these
two cases are related since $\theta_{B}$ is determined by the antenna
diameter $D$ ($\theta_B \approx \lambda/D$), and the shortest {\it
  possible} physically realizable value of $b_{min}$ is also
constrained by the dish diameter ($b_{min} > D$).  Interferometric
mosaicking can help in both cases, although in the second case
supplementary single-dish data will often be required depending on the
science goal and source morphology.

A related situation is that where the sources are compact by the above
criteria but distributed over a region comparable to or larger than
the antenna primary beam. Many interferometric surveys fall into this
category, such as the NRAO VLA Sky Survey \citep[NVSS:][]{condon1998}
and the more recent VLA Sky Survey \citep[VLASS:][]{lacy2020}.  These
wide-field imaging cases require techniques --- for instance, accurate
modelling of the primary beam and accounting for the W-term--- that
can also apply to the cases we are considering here. For clarity our
discussion will focus specifically on issues involved in imaging and
flux recovery for objects that are large in the sense indicated.  Note
that for geometrical reasons, the W-term tends not to be significant
for the types of compact configurations which are best for imaging
extended sources.

% original completion of above paragraph:
%The
%wide-field imaging techniques needed to deal with this third case are
%described in \citet[][{\bf THIS VOLUME}]{myers2015} and \citet[][{\bf
%    THIS VOLUME}]{rao2015}. While many wide-field considerations
%described--- for instance, A-projection, accurate modelling of the
%primary beam, and the W-term--- also apply to the cases we are
%considering here, for clarity our discussion will focus specifically
%on issues involved in imaging and flux recovery for large objects.
%Note that for geometrical reasons, the W-term tends not to be
%significant for the types of compact configurations which are best for
%imaging extended sources.

We can refine the criterion by which a source is deemed to be
``large'' and gain some intuition by considering a simple example:
measuring the visibility function on a single baseline of length $b$,
for a sky brightness component which has a Gaussian shape with FWHM $
= \theta_{src}$.  As a proxy for the quality of information on a given
angular scale, we consider the {\it ratio} of the measured visibility
to the total flux density of the Gaussian component $S_{src}$, which
can be shown to be:
\begin{equation}
\frac{V(b)}{S_{src}} = \frac{1}{1+(\theta_{src}/\theta_B)^2} \, {\rm Exp} \left(-4.71 \,
\frac{(b/D)^2}{1+(\theta_B/\theta_{src})^2} \right)
\end{equation}
Here we have assumed the antenna primary beam FWHM $\theta_B = 1.15
\times (D/\lambda)$.  This ratio is shown as a function of source size
for three represenatitive baseline lengths in Figure~\ref{gaussVis}.
Several facts are apparent. First, for a source as large as the
primary beam ($\theta_B = \theta_{src}$), even the shortest physically
realizable baseline ($b=D$) performs very poorly, recovering only
$5\%$ of the total flux density. For this reason, a typically quoted
``largest angular scale'' (LAS) that an interferometer can effectively
measure\footnote{The LAS a particular interferometer can accurately
  recover in a given observations depends on the $uv$-coverage. There
  is furthermore not a widely accepted quantitative criterion by which
  this LAS is judged.  Consequently there is some variation in the
  relationship between LAS and $b_{min}$ that different facilities
  quote. For example, the ALMA Technical Handbook gives $\theta_{LAS}
  = 0.6 \, \lambda / b_{min}$, and the VLA Observational Status
  Summary gives $\theta_{LAS} = 0.6 \lambda / b_{min} $ to $0.8
  \lambda / b_{min}$ for long tracks (half that for snapshots). If
  this level of discrepancy would substantially impact the science of
  a proposed project, careful simulations would be in order to
  determine the suitability of the proposed observations.} is
$\theta_{LAS} = \frac{1}{2} (\lambda/b_{min})$. If the $uv$-coverage
is poor, even this is an overestimate. Our single-baseline example
measurement would detect $\sim 40\%$ of the total flux density of a
Gaussian component of this size.  Second, it is apparent that the
decline in visibility with baseline length is very rapid. For
instance, a $b=2D$ basline--- which practically speaking is still a
very short baseline--- will recover only $4\%$ of the total flux
density for a Gaussian component with $\theta_{src} = \theta_{LAS} =
\frac{1}{2} (\lambda/b_{min})$. Finally, the {\it increase} in the
signal amplitude as the baseline gets shorter is (reciprocally) rapid:
a hypothetical $b=D/2$ baseline contains much higher quality
information about the large angular scales. The question is, can we
make use of information from such ostensibly unphysical baselines?  In
the next section we will see that an interferometer can in fact
recover much of the ``missing'', often critically important
short-spacing information, even without single-dish data.

% code for this figure is in mimas:/export/data_1/bmason/synthschool/
 \articlefigure{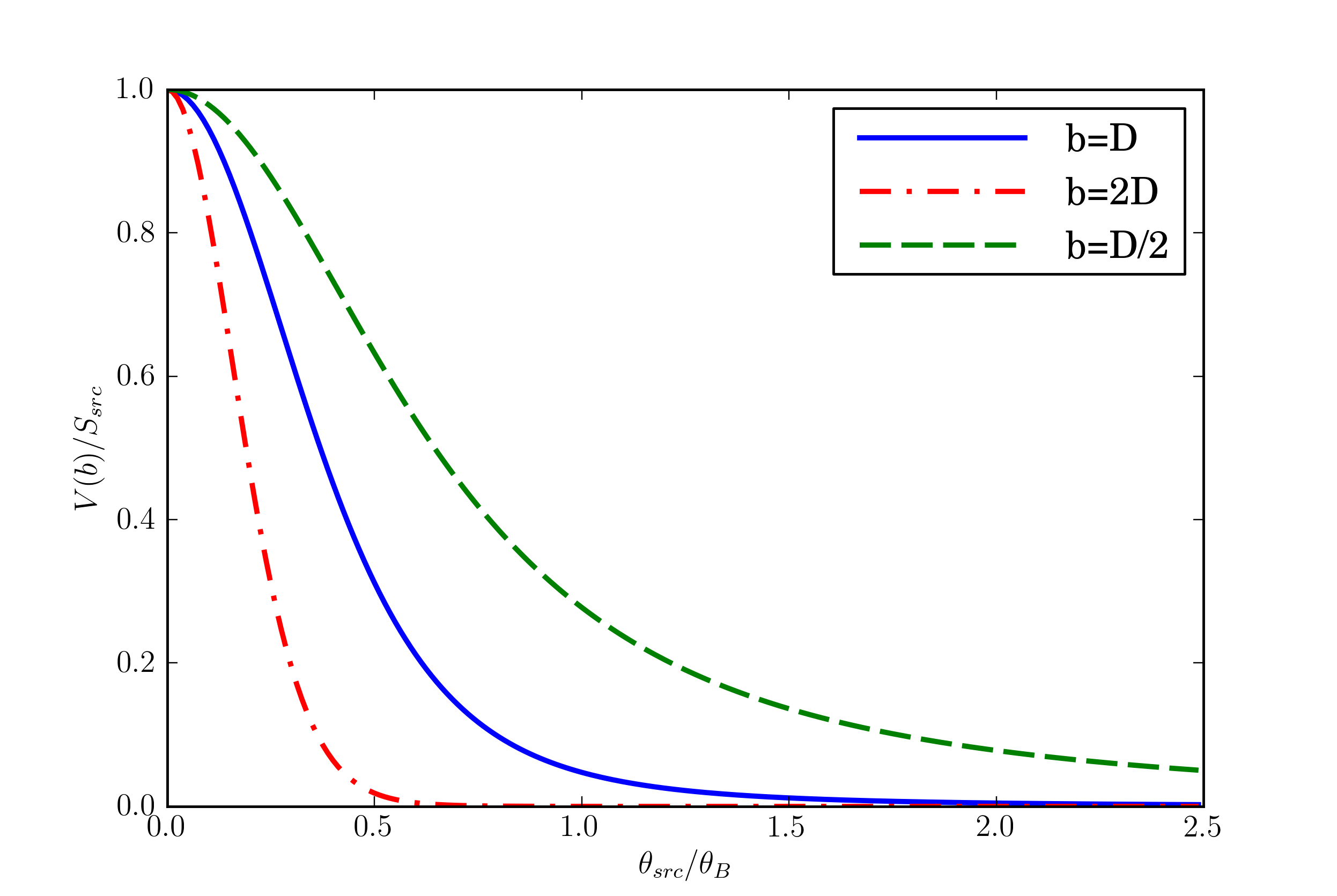}{gaussVis}{Fraction of total flux density
  retrieved by an interfereometer as a function of the ratio of the
  (Gaussian) source size to the antenna primary beam, for three
  different baseline lengths.}

A qualitative idea of the effect of missing short-spacings can be
obtained by considering Figure~\ref{braunwalterbos}. Imagining the
interferometer $uv$-coverage as a uniform plateau--- which has a
Fourier transform of a Sinc() function--- the short-spacing deficit
can be thought of as a uniform plateau of lesser extent {\it
  subtracted} from the $uv$-coverage.  This introduces a negative bowl
around the core of the point spread function (PSF), or dirty beam,
which in turn gives rise to negative bowls around regions of positive
emission in the dirty map.

\articlefigure{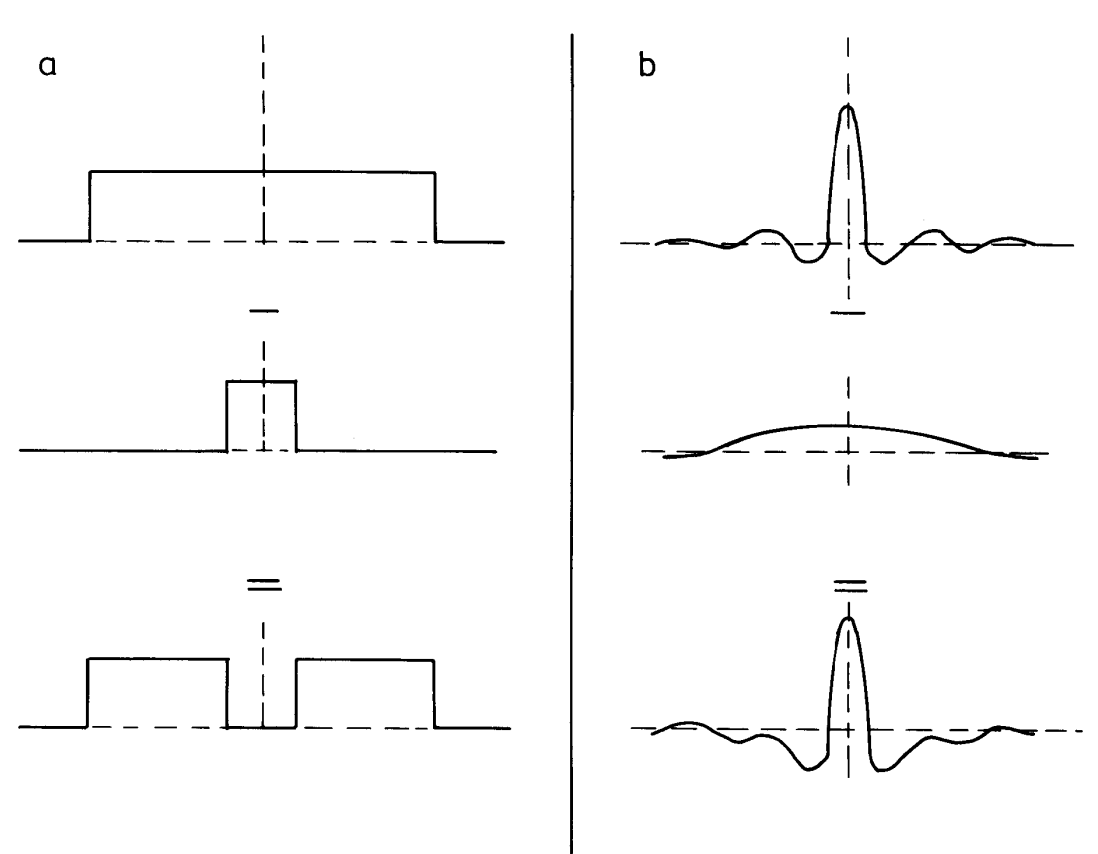}{braunwalterbos}{Conceptual
  illustration of the effect of missing short spacings on an
  interferometer's synthesized beam. \emph{Left (a): uv}- or
  aperture-plane coverage; \emph{Right (b):} resulting beam. \emph{Figures
    from Braun and Walterbos 1985}}

%\articlefigure{2012_antenna_0064.jpg}{apertureplane}{ALMA 7m antennas, illustrating b-D, b+D}

\subsection{The Ekers-Rots Theorem}

Begin by considering the measurement equation, which states the
relationship between a single visibility measurement $V(u,v)$ and the
sky intensity $I(\ell,m)$:
\begin{equation}
V(u,v) = \int \int \, d\ell \, dm \, A(\ell,m) I(\ell,m) \, e^{-2\pi i (u \ell + vm)}
\label{eq:measurement}
\end{equation}
Here $A(\ell,m)$ is the antenna primary beam (assumed identical for
all antennas); $I(\ell,m)$ is the sky intensity; $\ell$ and $m$ are
sky coordinates; $u$ and $v$ represent the baseline between antennas
in wavelengths; and $V(u,v)$ is the measured visibility value.  First,
note that $V(0,0)$ is the flux density which would be measured by a
single dish telescope with the same primary beam $A$; also known as
the ``zero-spacing'' flux\footnote{The term ``zero-spacing flux'' is
  sometimes loosely used to mean the total flux density of a given
  object, although if the source is comparable to or bigger than the
  primary beam $A(\ell,m)$, the total flux density is greater than the
  value $V(0,0)$ that a single-dish having that primary beam would
  measure.}. Most of the following also therefore applies to total
power measurements with single dish telescopes.

From the convolution theorem, Eq.~\ref{eq:measurement} can be written as
\begin{equation}
V(u,v) = \tilde{A}(u,v) \otimes \tilde{I}(u,v) 
\end{equation}
where $\tilde{I}(u,v)$ is the Fourier transform of the sky intensity
and $\tilde{A}(u,v)$ is the Fourier transform of the antenna primary
beam. From this it is clear that the antenna primary beam has the
effect of a point spread function in $uv$ space: a single visibility
measurement $V(u,v)$ has information not only from the Fourier
component $(u,v)$ of the ideal sky brightness $\tilde{I}(u,v)$, but
also from a range of nearby values determined by the Fourier transform
of the primary beam. 

% crap i need to remake this in omnigraffle
\articlefigure{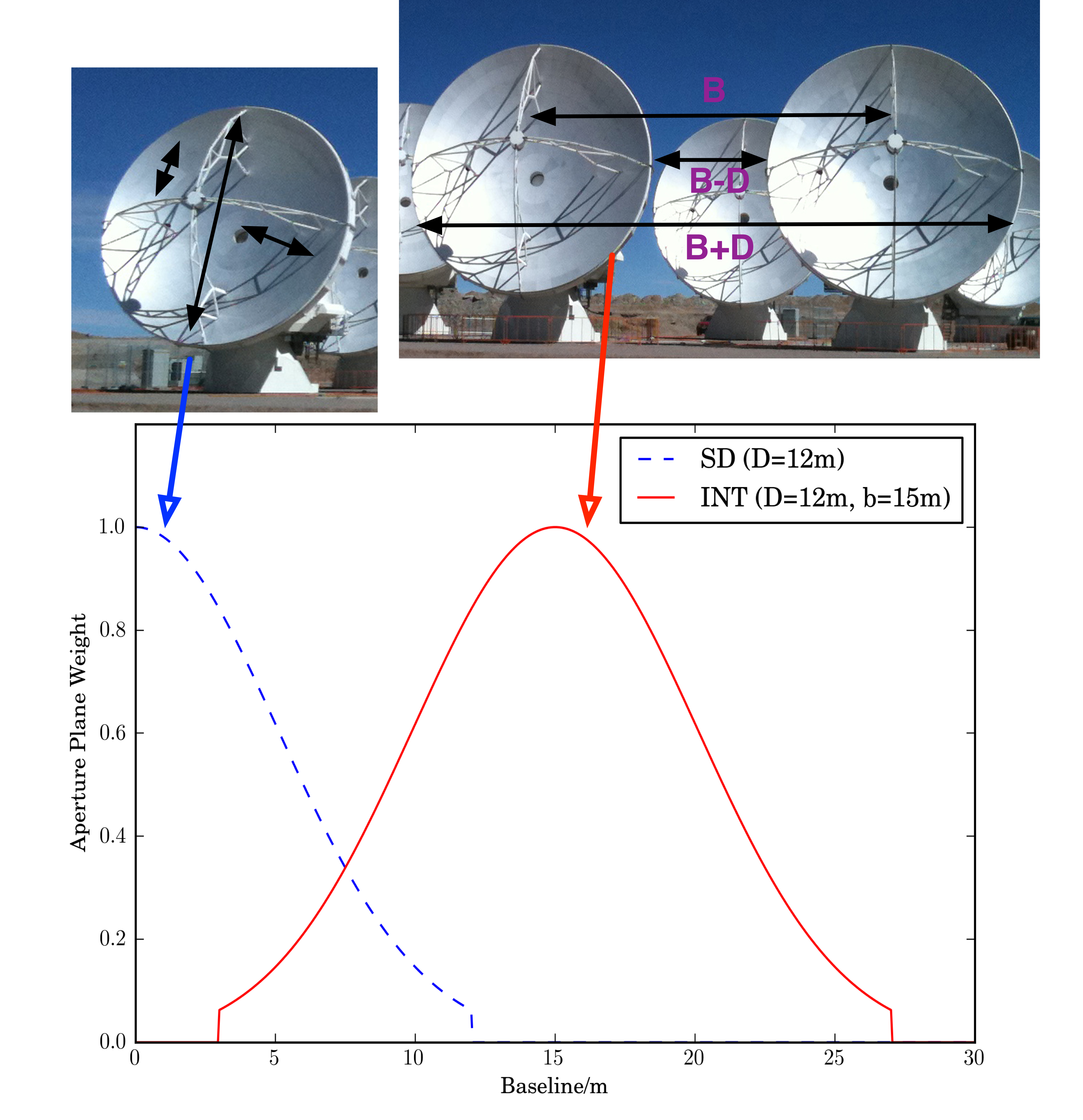}{fig:sdintuvweight}{Geometrical
  interpretation of the range of baselines measured by a single dish
  (blue, \emph{left}) and a single baseline interferometer (red,
  \emph{right}).}

Using the fact that the antenna primary beam $A(\ell,m)$ is the
magnitude squared of the Fourier transform of the aperture
illumination function $E(u,v)$ 
\citep{napier1999,hunter2016}
it can also be shown that
\[
\tilde{A}(u,v) = E(u,v) \otimes E(u,v)
\]
{\it i.e.}, the $uv$-plane ``point-spread function'' of an
interfometer is the auto-correlation function of the aperture
illumination function.  For a finite, circular aperture of diameter
$D$, $E(u,v)$ from $(u,v)=0$ to $\sqrt{u^2+v^2} = D/2\lambda$. It
follows that $\tilde{A}(u,v)$ will typically have support over a
region twice as large, \emph{i.e.}, from $(u,v)=(0,0)$ to
$\sqrt{u^2+v^2} = D/\lambda$.  \emph{Therefore, a given baseline $b$
  contains information not only about the spatial frequency}
$b/\lambda$\emph{, but about a whole range of spatial frequencies
  from} $(b-D)/\lambda$ \emph{to} $(b+D)/\lambda$.

This result is illustrated in Figure~\ref{fig:sdintuvweight} : a
single visibility measurement on a baseline of (nominal)
length $b$ contains information about a range of baseline lengths,
from $b-D$ to $b+D$. Similarly, a measurement with a single-dish
telescope can be thought of as being a sum of measurements with
``baselines'' ranging from $D$ to $0$. Geometrically it is clear that
the single dish measurement has many of the very short spacings and
few ``baselines'' of length $\sim D$; the single dish correspondingly
provides greater sensitivity or weight at the shortest spatial
frequencies and less at the higher spatial frequencies.  Similarly the
interferometer provides the highest sensitivity at the spatial
frequency $b/\lambda$.

\articlefigure{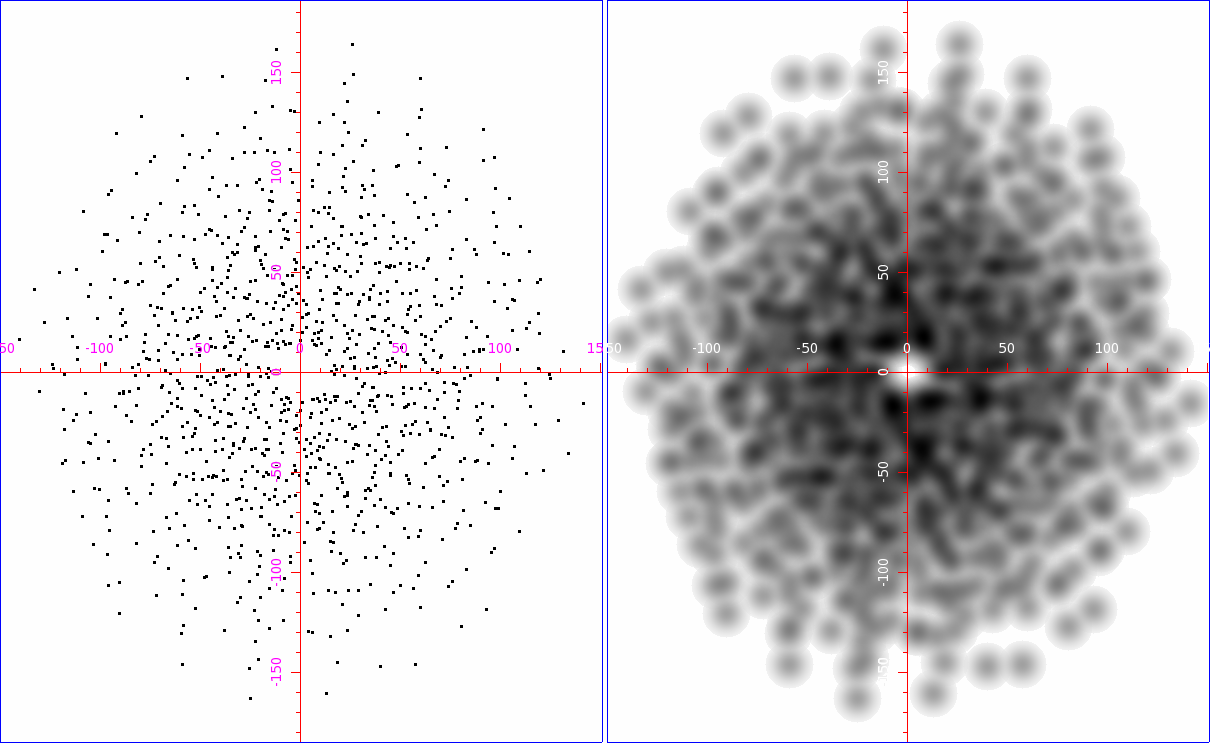}{fig:almauvblur}{Nominal aperture plane
  coverage of a single snapshot of a compact ALMA configuration
  (\emph{left}). The effective aperture plane coverage is shown on the
  \emph{right}, which is simply the convolution of the points in the
  left panel with antenna (aperture-plane) primary beam. Note that the
  units of aperture plane coordinates are \emph{distance} units (in
  this case, meters).  This is a convenient representation of an
  interferometer's spatial frequency response when, as is often the
  case by design, the antenna illumination pattern is approximately
  wavelength-independent.}

%A representative illumination function is shown
%in Figure~\ref{fig:illum}, along with its aperture plane
%auto-correlation.  

%A single-dish measurement can be considered to be the $(u,v)=(0,0)$
%case of this expression, with $V(0,0)$ representing the measured mean
%level of the sky at the (implicitly specified) location.

%\[
%V(u,v) =  FT[A(\ell,m) I(\ell,m)]
%\]

The effect of the blurring effect of the primary beam on an
interferometer's $uv$-coverage can be seen in
Fig.~\ref{fig:almauvblur}.  This is particularly important for the
problem of imaging large objects: the ``zero-spacing-hole'' has
shrunk, and the shortest interferometer baseline $b \sim D$ is seen to
contain information almost all the way down to $(u,v)=0$. The very
inner-most (lowest spatial frequency) points will have poor
signal-to-noise due to their low weight; these correspond to spatial
frequencies of $\sim (3{\rm m})/\lambda$ in
Fig.~\ref{fig:sdintuvweight}. The other effect of the primary beam
convolution has been to fill in most of the unsampled holes in the
aperture plane, at least for this (quite compact) array
configuration. If this information can be used it has the potential to
stabilize nonlinear deconvolution algorithms like CLEAN
\citep{cornwell1999}.

The problem is that with only a single measurement $V(u,v)$, it is
impossible to separately estimate the contributions of the individual
Fourier components of the sky brightness. \citet{ekers1979} first
showed how this limitation can be overcome for an interferometer.  By
continuously scanning the antennas' pointing positions, making many
measurements at different pointing centers on the sky $(\ell',m')$,
one can tabulate (sample) the function $V(u,v,\ell',m')$. Fourier
transforming this function with respect to $\ell '$ and $m'$ yields an
estimate of the sky Fourier transform which has a Fourier resolution
determined by the {\it size of the area scanned over}. If this area is
larger than the primary beam, this is a higher Fourier resolution
estimate of the sky brightness. It will also allow estimation of sky
components with spatial frequencies all the way down to
$(b_{min}-D)/\lambda$, {\it i.e.}, less than the spatial frequency
given by the shortest physical baseline $b_{min}/\lambda$. A
completely analagous argument applies to measurements made by scanning
a single dish telescope.

The scheme described by Ekers \& Rots gives powerful conceptual
insight but suffers from several shortcomings as a practical solution
to the short spacing problem.  A variety of practical implementations
are described in \S~\ref{sec:mosaicking}. Among the shortcomings of
Ekers-Rots is the fact that what is usually of interest is not a
higher resolution estimate of the sky's Fourier transform, but rather,
an improved image of the sky--- and in particular, an image which has
not been convolved with the interferometer's synthesized
beam. Consequently most practical algorithms which make use of the
Ekers-Rots information do so within the context of a deconvolution
operation.  Additionally, most interferometers do not retain fixed
baselines $(u,v)$ with time; rather, the baslines (projected on the
sky) change as time progresses. Finally, continuously-scanned
interferometric observing has not been commonly available on major
radio interferometers until relatively recently.

%The next sub-section will discuss the sampling requirements for mosaic
%observations in more detail.

%Another practical problem is that for most interferometers, the
%projected baseline for a given pair of antennas changes with time and
%pointing position.

\subsection{Mosaic Observing Strategy \& Fourier Space Sampling}
\label{sec:sampling}

Although the argument of Ekers \& Rots assumes the interferometer is
continuously scanned across the sky, it turns out that continuous
scanning is not in fact required.  \citet{cornwell1988} showed that
the full Ekers-Rots information can be obtained provided only that the
sky is Nyquist sampled with respect to the antenna primary
beam. Cornwell also presented one of the first practical
implementations of a deconvolution algorithm which has the benefit of
the short-spacing, Ekers-Rots information. This algorithm will be
discussed in \S~\ref{sec:cornwell}.

The Nyquist sampling requirement can be understood as follows.
Imagine a single-baseline interferometer with a fixed baseline
$(u,v)$, constructing a rectangular mosaic of pointings spaced by
$\Delta$ in each of two orthogonal directions on the sky. This will
give rise to a series of aliases spaced by $1/\Delta$ in the aperture
plane. Each measurement, and each alias, is in fact a measurement over
a circular patch\footnote{Assuming that the antennas are circular, as
  is commonly but not always the case.} of $uv$-space of diameter
$2D/\lambda$. Therefore spacing the interferometer pointings $\Delta
\le \lambda/2D \sim \theta_B/2.3$ suffices to prevent the aliases
from overlapping the ``real'' data.

Because the Nyquist-Shannon theorem strictly applies only in one
dimension, it is in fact possible to do slightly better than this. A
generalization of the Nyquist-Shannon theorem \citep{petersen1962} can
be used to show that in two dimensions the optimal sampling strategy
for circularly band-limited functions is a {\it hexagonally closed
  packed} lattice. This gives the maximally close-packed arrangement
of aliases in Fourier space, and therefore a somewhat more sparse set
of telescope pointings in real space.  For the case of mosaicking with
antennas of diameter $D$ it corresponds to interferometer pointing
arranged in equilateral triangles with centers spaced by
$\lambda/\sqrt{3} D \sim \theta_B/2$. A hexagonal pattern will require
$15\%$ fewer pointings than a rectangular mosaic covering the same
region.

For imaging extended emission with an interferometer the recommended
practice is to use a hexagonal mosaic with pointing centers spaced by
$\lambda/\sqrt{3} D \sim \theta_B/2$ or even closer. If the
measurement bandwidth is significant then the shortest wavelength of
interest $\lambda_{min}$ should be used to establish the sampling
requirement.  In practice, the amount of aliasing caused by slightly
undersampling the sky is greatly reduced by the effects of the
aperture illumination taper. Therefore for surveying relatively
compact sources over large regions considerably more sparse
samplings--- $0.7 \, \theta_B$ or even $ 0.8 \, \theta_B$--- are often
used to good effect.  Note that the anti-aliasing sampling
requirements $\lambda/2D$ (for a square mosaic) and $\lambda/\sqrt{3}
D$ (for a hexagonal mosaic) are exact and do not require use of an
approximate Gaussian beam width.  As pointed out by
\cite{cornwell1988}, all of these sampling arguments also apply to
single-dish mapping.

An approach which has gained wider use recently due to advances in
computing and storage speed is {\it on-the-fly} (or OTF) mosaicking
\citep[e.g.][]{at20g,lacy2020}.  The OTF observing technique is widespread in
the single-dish community; \citet{mangum2007} present an excellent
review of the details involved in using OTF in a single dish context.
In interferometric OTF mosaicking--- just as for single-dish OTF---
the antenna pointing positions are continuously and synchronously
scanned over the region of astronomical interest on the sky while the
visibility data are recorded, usually at a considerably higher data
rate than would be necessary for a conventional mosaic.  This is also,
of course, the observing scheme considered by \citet{ekers1979}.  OTF
mosaicking will often provide higher observing efficiency ({\it i.e.},
lower observing overheads), and can be advantageous in situations
where the goal is to map large areas quickly.

%Interferometric OTF
%mapping is also discussed in \citet[][{\bf THIS VOLUME}]{myers2015}.

\section{Interferometric Mosaicking Algorithms}
\label{sec:mosaicking}

In the following sub-sections, I describe the algorithms most commonly
used to make images from mosaicked interferometer data.

\subsection{Linear Combination of Separately Deconvolved Images}
\label{sec:linearmosaic}

The most straightforward approach to making an image from mosaicked
interferometer data is the so-called {\it linear mosaic}. In a linear
mosaic the individual interferometer pointings are deconvolved
separately and the resulting deconvolved maps are appropriately
weighted and added together to make a final mosaic image. Explicitly,
the linear mosaic image $I_{LM}$ formed out of the individual-pointing
deconvolved images $I_p$ is:
\begin{equation}
I_{LM}({\bf x}) = \frac{\sum_p \, A({\bf x}-{\bf x}_p) I_p({\bf x})/\sigma_p^2}{\sum_p \, A^2({\bf x}-{\bf x}_p)/\sigma_p^2}
\end{equation}
Here $\sigma_p$ is an estimated or measured noise level for pointing
$p$ which is needed to weight the data for optimal sensitivity in
beam-overlap regions.  For the single-field case this reduces to
\[
I_{LM}({\bf x}) = \frac{I_p({\bf x}) }{A({\bf x}-{\bf x}_p)}
\]
which is simply the deconvolved image corrected for the primary beam
attenuation.

Linear mosaicking is not usually recommended for imaging extended
objects. It suffers from two significant disadvantages for this
application, both related to the critical deconvolution step. First,
because the visibility data from the individual pointings are never
used together, the deconvolution algorithm does not have access to the
Ekers-Rots information. Therefore the Fourier resolution and
short-spacing information in the deconvolved image will be no better
than for a single field. Second, the signal to noise ratio and
$uv$~coverage in the individual (dirty) maps used by the deconvolution
is worse than in the joint map, particularly in regions of significant
overlap between adjacent pointings. This also limits the effectiveness
of the deconvolution algorithm, {\it e.g.}, how deeply it is feasible
to clean in order to accurately deconvolve diffuse structures. These
problems are exacerbated due to the fact that the deconvolutions used
for interferometric imaging are necessarily nonlinear operations.

On the other hand linear mosaicking does offer the possibility of
carefully fine-tuning the calibration and deconvolution parameters for
each individual pointing.  This can be useful when there are
significant time and sky position variable effects (such as the
ionosphere), or when the field is crowded or confused. For these
reasons linear mosaicking is often used at low frequencies ($<1 \,
{\rm GHz}$).

% a good reference -
% http://www.aoc.nrao.edu/~rurvashi/ImagingAlgorithmsInCasa/ImagingAlgorithmsInCasa.html

Because it is simply a linear combination of all of the visibilities,
a linear mosaic of individual-pointing {\it dirty images} does in fact
contain the full Ekers-Rots information. Since it is convolved with
the interferometer synthesized beam--- which moreover will vary from
pointing to pointing, and thus over the mosaicked dirty image--- the
joint dirty image mosaic is not useful for most applications.  This
observation does, however, motivate one of the methods of {\it joint
  deconvolution}  presented in the next section.

Linear mosaicking is available via the CASA toolkit image toolkit
method  {\tt linearmosaic}; there is also a prototype implementation
in {\tt tclean} accessed by setting {\tt gridder = `imagemosaic'}. It is available
in AIPS through the {\tt FLATN} task.

\subsection{Joint Deconvolution}
\label{sec:jointdeconv}

% should i call this ``imaging'' 
% make general comment - we use ``deconvolution'' in a general
%  sense of forming an estimate of the true sky brightness 
%  distribution / reducing the effect of finite uv sampling.
% not a technical ``convolution'' which entails a position independent
%  PSF by definition.

In order for the deconvolved interferometer image to possess accurate
information at spatial frequencies shorter than that of the shortest
baseline, it is necessary for the deconvolution algorithm to make use
of the visibilities from all of the interferometer pointings
\citep[see, {\it e.g.},][]{cornwell1988,cornwell1993,sault1996}.  I 
will describe three approaches that have been widely used: one that
combines the visibility data during deconvolution
(\S~\ref{sec:cornwell}), and two that combine the visibility data from
different pointings before the deconvolution (\S~\ref{sec:sault} and
\S~\ref{sec:widefield}).  Note that I use the term ``deconvolution''
in a broader sense--- common in the synthesis imaging community--- of
estimating the true sky brightness from the incompletely sampled data,
rather than the stricter mathematical sense of removing the effects of
a spatially invariant point spread function.

\subsubsection{``Nonlinear'' Joint Deconvolution \& Maximum Entropy}
\label{sec:cornwell}

CLEAN is a procedure or recipe which decades of use by the community
has shown to be quite effective for deconvolution of interferometric
data. Its nature as a procedure, however, can make its results and
limitations difficult to understand. An alternative approach is to
determine the true sky brightness as the solution to a more
mathematically simple and well-defined problem.  One could, for
instance, regard the true (deconvolved) sky as represented by a set of
pixel values $I_j$, all of which are parameters which are varied in a
$\chi^2$ fit to the measured visibility data $V$:
\begin{equation}
\chi^2 = \sum_p \sum_i \frac{|V_{p,i} - V^M_{p,i}|^2}{\sigma_{p,i}^2}
\label{eq:chi}
\end{equation}
Here the model visibility values $V^M_{p,i}$ for each pointing $p$ are
obtained by Fourier transforming the sky model $I_j$ using the
measurement equation.  Due to the finite sampling of $uv$ space, this
minimization is not generally a well-posed or stable problem: this is
the ``ghost distributions'' problem discussed in
\citet{cornwell1999}.

%Another way to view the problem is
%that the {\bf R} matrix of the imaging problem--- viewed as a linear
%least squares problem--- is ill-conditioned \citep[][{\bf THIS VOLUME}]{myers2015}.

\citet{cornwell1988} introduced a practical algorithm for deconvolving
mosaicked interferometer data which, instead, aims to maximize the
so-called {\it Entropy}
\begin{equation}
H = - \sum_j I_j \, {\rm ln} \left(\frac{I_j}{M_j e}\right)
\label{eq:entropy}
\end{equation}
subject to the constraint that $\chi^2$--- Eq.~\ref{eq:chi}--- is
close to its expected value. In Eq.~\ref{eq:entropy}, $M_j$ is a
``default image'' which is an input to the deconvolution process. In
the absence of other information it is typically taken to be a
constant.  The $H$ term serves to regularize the otherwise ill-posed
inversion problem and thereby select one of the infinitely many
surface brightness distributions which are consistent with the
measured visibilities.  \citet{cornwell1988} called this method
``Nonlinear Mosaicking''. It is commonly referred to in the synthesis
imaging literature as the Maximum Entropy Method (MEM). A general
review of the MEM technique is given by \citet{narayan1986}.

Maximizing the entropy term has the effect of compressing the range of
pixel values (relative to the default image) in the solution. This
tends therefore to produce smoother images than CLEAN, and is
naturally suited to imaging extended emission.  The images MEM
produces are biased for two reasons: the pixel values $I_j$ in the
solution must be positive; and the effect of the entropy term $H$ is
typically to move the solution away from solutions which strictly
minimize $\chi^2$ alone. Means of minimizing this bias will be
discussed more in \S~\ref{sec:clean}. One attractive property of MEM
is that it provides a straightforward method to provide single dish
information to the joint, interferometric deconvolution: the single
dish data can be used as the default image $M_j$
(\S~\ref{sec:combining}).

A version of the MEM mosaic deconvolver is implemented in AIPS as the
tasks {\tt UTESS} and {\tt VTESS}, and is available in the CASA toolkit.

\subsubsection{Deconvolution of Combined Dirty Maps}
\label{sec:sault}

Another approach to imaging mosaicked interferometer data takes as its starting
point the linear mosaic of dirty maps from all pointings:
\begin{equation}
I^D_{joint}({\bf x}) = W({\bf x}) \, \frac{\sum_p \, A({\bf
    x}-{\bf x}_p) I^D_{p}({\bf x})/\sigma_p^2}{\sum_p \, A^2({\bf
    x}-{\bf x}_p)/\sigma_p^2}
\label{eq:dirtyjoint}
\end{equation}
$W({\bf x})$ here is an apodization function used to suppress noise
artifacts at the edge of the mosaic.  As previously noted this joint
dirty map contains all of the information that the visibilities
together do, but not in a very convenient form since it is convolved
with a messy point spread function. In fact for an image formed
according to Eq.~\ref{eq:dirtyjoint}, the PSF will typically be
spatially variable since the $uv$ coverage and noise generally change
from one interferometer pointing to the next. This approach is often
referred to simply as ``joint deconvolution''.

Given a method of calculating the PSF this $I^D_{joint}$ can be
deconvolved by the usual means, {\it e.g.} CLEAN or
MEM. \citet{cornwell1993} studied a method employing an approximate
PSF evaluated at the center of the mosaic; CASA's {\tt tclean} task,
invoked with {\tt gridder='standard'}, uses a variant of this
algorithm.  \citet{sault1996} present an exact expression for the
position-dependent PSF of the map in Eq.~\ref{eq:dirtyjoint}; they
also present an image projection designed to minimize w-term
distortions.  MIRIAD implements this form of joint deconvolution using
in tasks {\tt MOSSDI} (which uses CLEAN for the deconvolution) and {\tt
  MOSMEM} (which uses MEM instead). Use of MEM for the deconvolution of the joint
dirty map (Eq.~\ref{eq:dirtyjoint}) is very similar to the strategy
outlined in \S~\ref{sec:cornwell}, but with the $\chi^2$ of the sky
model with respect to the data evaluated in the image rather than
visibility domain.

This approach to joint deconvolution has been widely used and is a
good option for imaging significantly extended objects with an
interferometer.

\subsubsection{Wide-field Imaging}
\label{sec:widefield}

A more modern approach to imaging mosaicked interferometer data is to
reference all pointings' data to a single phase center and grid them
onto a common $uv$-plane. This results in a single PSF with improved
$uv$-coverage for the whole mosaic and an optimal-sensitivity
weighting of the data. This approach is described in detail in
\citep{myers2003} but a brief outline is as follows. First, we can
write the measurement equation for a given visibility ${\bf u}_k =
(u_k,v_k)$, for a mosaic pointing at ${\bf x}_p = (\ell_p,m_p)$, and
with an explicit phase tracking center ${\bf x}_{\phi,p}$, as:
\begin{equation}
V_p({\bf u}_k) = \int \int \, {\bf d^2x} \, A({\bf x}-{\bf x}_p) I({\bf x}) \,
e^{-2\pi i {\bf u}_k \cdot ({\bf x} - {\bf x}_{\phi})}
\label{eq:measurement2}
\end{equation}
Using the shift theorem this can be re-expressed as an integral in the aperture plane as
\begin{equation}
V_p({\bf u}_k) = e^{2\pi i ( {\bf u}_k \cdot {\bf x}_{\phi,p} ) } \,
\int \int \, {\bf d^2u} \, \tilde{A}({\bf u}_k-{\bf u}) \tilde{I}({\bf u}) \,
e^{2\pi i ({\bf u} - {\bf u}_k) \cdot {\bf x}_p} .
\end{equation}
This relationship can, in essence, be inverted to obtain a set of
estimators $\tilde{I}_j = \tilde{I} ({\bf u}_j)$ of the Fourier
transform of the sky brightness on a grid of points ${\bf u}_j$. The
cell size of the grid ${\bf u}_j$ is determined by the angular extent
of the mosaic on the sky and the estimators themselves are simply
linear combinations of the original visibility data with appropriate
phase gradients and normalizations. To the extent that the aperture
plane illumination $\tilde{A}({\bf u}_k-{\bf u})$ is correct, the
``blurring'' effect of the interferometer primary beam will have been
removed from the $\tilde{I}_J$, which will have a resolution
determined by the mosaic size. These visibility estimators can be
imaged by standard techniques such as CLEAN.

This approach makes optimal use of the whole range of aperture plane
information inherent in the data and provides a natural basis on which
to incorporate wide-field effects such as W-projection and
A-projection. It is also well-suited to jointly imaging data from a
heterogeneous interferometer, \emph{i.e.}, one in which the antenna
sizes are not identical.  One drawback to this approach is that it
relies upon an accurate model of the antenna primary beam.  It can
also be computationally expensive, though it is faster when there are
many pointing centers in the mosaic, such as there are for an OTF
mosaic.  It is generally the preferred approach for mosaicking
extended structures when using CLEAN in CASA. The {\tt tclean} task
uses this algorithm when {\tt gridder='mosaic'} is set in its
invocation.

% eqs see slide 45 of myers talk.
%  cite myers 2003
%  maybe sanjay 2008 A&A paper
%

\subsection{Practical Challenges in Deconvolution}
\label{sec:clean}

% can i use ``Imaging'' for ``Deconvolution''? 

Even with the benefit of the information provided by the joint
mosaic pointings' visibilities, imaging extended emission with an
interferometer can be challenging.  It is therefore important to
understand the characteristics of the deconvolution algorithm used.
Here I summarize some relevant, key properties of the two dominant
approaches used by the synthesis imaging community: CLEAN and Maximum
Entropy.

Since CLEAN represents the sky as a sum of {\it unresolved} signals
(or ``point sources''), it can take many iterations to construct a
model of a significantly extended source. In general, it is advisable
to clean interactively; to carefully define the regions used for
cleaning so as to contain only ostensibly real and believable signal;
and to iteratively inspect and refine these regions through the
cleaning process ({\it e.g.}, after each major cycle). For imaging
extended emission, the CLEAN threshold chosen can have a significant
effect on the resulting image. For instance, a $3\sigma$ threshold
will leave a plateau of un-deconvolved signal of approximately this
amplitude (with the corresponding ``short spacing bowl'' around
it). This effect can be mitigated by cleaning deeply, {\it e.g.}, down
to a $1.5\sigma$ threshold. Such deep cleaning will result in some
spurious noise components in the CLEAN model. This is often acceptable
since the science image is usually taken to be the sum of the
(CLEAN-beam convolved) model and the residuals, and to first order the
deep cleaning will simply have re-partitioned the noise between the
model and the residuals.  It is potentially more problematic if the
CLEAN model is to be used for self calibration. Such deep cleaning can
also be very time-consuming for spectral line cubes with many
channels.

There are two other biases worth bearing in mind when using CLEAN to
image extended objects.  One is the flux bias resulting from the
mismatch between the clean and dirty beam areas, discussed by
\citet{jorsater1995} and \citet{walter2008}. These authors give an
analytic correction for the effect, and note that it is also mitigated
by deeper cleaning. Another is the so-called ``CLEAN bias'' discussed
in, \emph{e.g.}, \citet{condon1998}. The CLEAN bias results from the
constructive interference of the sidelobes of the synthesized beam.
Particularly for diffuse, extended sources, these sidelobes can then
be brighter in the dirty map than the apparent signal in the main lobe
of the PSF. This biases the recovered flux densities low and is
mitigated by careful masking and better $uv$ coverage. All of that
said, CLEAN is the algorithm most commonly used for synthesis imaging,
and when carefully applied will generally yield excellent results if
the data are of sufficient quality.

A variety of techniques have been developed to improve the performance
of CLEAN for extended sources. These include Multi-Scale CLEAN
\citep{cornwell2008} and ASP-clean \citep{bhatnagar2004}, and involve
using CLEAN components of varying sizes.  These algorithms can improve
the stability and convergence speed when imaging extended
objects. Multi-scale CLEAN is available within CASA's {\tt
  tclean} task by setting {\tt deconvolver = multiscale}.

Maximum Entropy Deconvolution (MEM) is naturally well-suited to image
extended emission. By definition it is a biased estimator of the true
sky image since the effect of including of an entropy term is to move
the solution away from the minimum of $\chi^2$
(\S~\ref{sec:cornwell}). For SNR $>>1$ this bias is negligible. The
bias is also reduced by using an appropriate default image, {\it
  e.g.}, a single-dish image. The resolution of MEM-reconstructed
images is also known to vary with SNR; the impact of this can be
reduced by convolving the MEM image with a nominal synthesized beam,
much as is done with a CLEAN model. Finally, MEM has great difficulty
with point sources embedded in diffuse emission. In this circumstance
it is best to remove the point source by other means (such as CLEAN)
prior to MEM deconvolution.

Finally, different implementations of these deconvolution algorithms
often contain subtle (or not subtle) differences, and the performance
of algorithms can vary between implementations. If the application is
demanding--- and imaging diffuse emission often is--- it can be
helpful to consult or collaborate with an expert.

\section{Combining Interferometeric and Single-Dish Data}
\label{sec:combining}

Often when the source of interest is comparable in size to the
interferometer primary beam the best course of action is to obtain
single-dish data and to combine the two datasets.  A variety of
techniques has been developed to do this. An excellent review and
comparison of techniques used to combine interferometric and single
dish data is given by \citet{stanimirovic2002}. Here--- because it is
straightforward, widely used, and robust--- I will focus mainly on a
technique to linearly combine the single dish and interferometer data
in $uv$ space known as ``feathering''.  \citet{cotton2017} discusses 
the technique in  detail.

To understand feathering, it is useful  first to consider the nature of
the low spatial frequency components in an interferometer map which
has been deconvolved using CLEAN. Although the interferometer does not
accurately measure these spatially large signals there is not usually
an explicit constraint on the total flux density in the CLEAN model: for
instance, it is not generally zero. Therefore CLEAN will
\emph{extrapolate} the values of spatial frequencies
smaller than the smallest measured.  Often this extrapolation will be
very noisy.  It is desirable to down-weight this low-quality
information where better quality (single dish) information is
available.

To feather single dish and interferometric maps together, the
individual datasets are first deconvolved separately to remove the
effects of their respective point spread function.  Each deconvolved
map is then Fourier transformed and the interferometer map is
high-pass filtered--- with a characteristic scale determined by the
diameter $D_{SD}$ of the single dish telescope--- to eliminate the
poorly determined large-scale information in it. The maps are then
added together and inverse Fourier transformed to obtain a single
image with information from both datasets.  In more detail, the
sequence of steps typically required is as follows:
\begin{enumerate}
\item Prepare input images.
\begin{enumerate}
\item Deconvolve the synthesized beam from the interferometric dirty
  map, {\it e.g.} using CLEAN.  The resulting model is usually then
  re-convolved with a nominal ``restoring'' beam (of FWHM
  $\theta_{B,int}$) to obtain the deconvolved interferometric map
  $I_{int}$ in units of Janskys per (restoring) beam. This nominal
  beam is almost always taken to be a Gaussian. 
\item Deconvolve the antenna beam from the single dish map and
  re-convolve it with a nominal restoring beam ($\theta_{B,SD}$) to
  obtain the deconvolved single dish map $I_{SD}$ in Janskys per
  (restoring) beam. Since the single dish data in principle sample all
  spatial frequencies out to the maximum ($D_{SD}/\lambda$), simple
  linear methods can sometimes be used instead of the non-linear
  methods required by the interferometric imaging problem. If the
  single-dish beam is simple enough to be well-approximated by a
  Gaussian, this step can be omitted. If there is considerable overlap
  between the interferometer and the single dish in $uv$-space, it may
  be desirable to use a restoring beam larger than the antenna primary
  beam to give relatively greater weight to the interferometer data at the higher
  spatial frequencies.
\item Various implementation-dependent clerical steps may also be
  required. For example: ensuring the map sizes, registrations, and
  pixellizations are appropiate; correcting for the primary beams;
  re-ordering axes; padding and apodizing maps as needed to avoid edge
  effects; and putting restoring beam information in image headers.
\end{enumerate}
\item Place the deconvolved maps on a common surface brightness scale.
  This can be done, for example, by multiplying $I_{SD}$ by
  $(\theta_{B,int}/\theta_{B,SD})^2$, or by converting both maps into
  instrument-independent surface brightness units such as Janskys per
  Steradian. 
\item Fourier transform the maps to obtain $\tilde{I}_{SD}(u,v)$ and
  $\tilde{I}_{int}(u,v)$.
\begin{enumerate}
\item It is often useful at this stage to correct any (hopefully slight) errors in
  the relative calibration of the single-dish and interferometric
  maps, \emph{e.g.}, due to the flux scale assumed. This can be done
  by deriving a scale factor $f_{cal}$ from comparing the Fourier
  transforms of $I_{SD}$ and $I_{int}$ in the range of $uv$ values
  well-measured by each, properly accounting for the restoring beam of
  each.
\item ``feather'' the images together in the Fourier domain using a taper function $T(u,v) =
  1-\tilde{B}_{SD}$ as
\begin{equation}
\tilde{I}_{combined}(u,v) = \tilde{I}_{SD}(u,v) + \tilde{I}_{int}(u,v) \times (1- \tilde{B}_{SD})
\end{equation}
where $\tilde{B}_{SD}$ is the Fourier transform of the single dish
restoring beam normalized to have a peak of one.  An example
interferometer re-weighting function suitable for combining GBT and
VLA C-configuration data is shown in Figure~\ref{gbtVlaFeather}. The
taper function has the effect of emphasizing each dataset where it
provides the best information while minimizing excessive noise that
will be introduced by over-weighting data in poorly measured regions
of the $uv$ plane.
\end{enumerate}
\item Inverse Fourier transform to obtain the image $I_{combined}$,
  containing information from both the single dish and the
  interferometer.
\end{enumerate}
This algorithm is implemented in CASA as {\tt feather}\footnote{A word
  of caution: the {\tt sdfactor} keyword of CASA's {\tt feather} task
  is not a weight, it is a direct scaling factor applied to the single
  dish data. As such, values other than the default ($1.0$) should be
  set with caution, and a clear and quantitative physical
  motivation.}, in AIPS as {\tt IMERG}, and in MIRIAD as {\tt
  IMMERGE}.  These implementations differ in the details of how the
restoring beams and feathering weights are defined and handled, but
typically take the deconvolved interferometer and single dish maps and
their beams $\theta_{SD}$, $\theta_{int}$ as input and perform steps 2
- 4 of the above.  They also assume that the restoring beams are
Gaussian. In particular, as described in step 1(b) above, if the
single-dish beam significantly deviates from a Gaussian shape--- {\it
  e.g.} due to sidelobes or ellipticity--- then its effects should be
removed from the map by deconvolution \citep[see, {\it
    e.g.},][]{weiss2001}.  Two of many examples of the application of
the feathering technique in the literature are \citet{vogel1984} and
\citet{dubner1998}.

\articlefigure{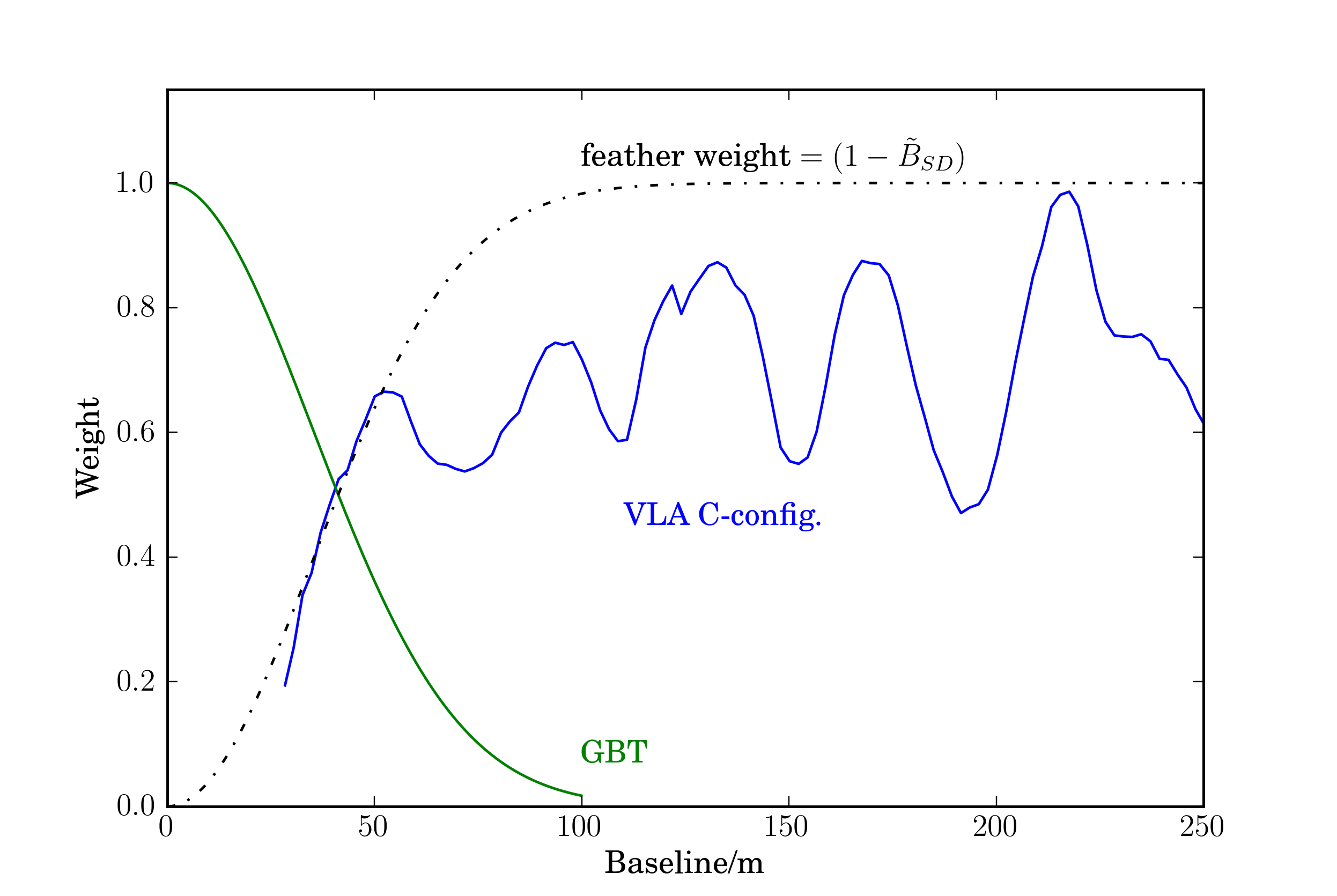}{gbtVlaFeather}{Relative
  aperture-plane sensitivities of GBT and the C-configuration of the
  VLA and an appropriate feathering weight which could be used to
  combine the two.}

For feathering to work well there should be region of the $uv$ plane
measured by both the interferometer and the single dish.  A commonly
adopted criterion is for the diameter of the single dish $D_{SD}$ to
be $1.5$ to $2$ times larger than the minimum baseline $b_{min}$ in
the interferometric array. By this criterion, for instance, the
100-meter GBT is well-suited to provide short-spacing data for EVLA C-
or D-configuration data.  This criterion also implies that an
interferometeric array with antennas of a single diameter cannot even
in principle provide high-quality total power data for itself by
measuring auto-correlations or outfitting some of the antennas with
total power receivers.

Another approach to combining single dish and interferometric data is
to do so before or during the deconvolution.  This has the advantage
of providing considerably more information to the nonlinear
interferometric deconvolution. One method, mentioned previously in
\S~\ref{sec:cornwell}, is to use the single-dish data as the ``default
image'' in a Maximum Entropy deconvolution of the interferometric
image. Another method is to directly form a linear combination of the
single-dish and interferometer ``dirty'' images, pre-deconvolution. In
this case the effective beam is the corresponding linear combination
of the respective single-dish and interferometer point spread
functions; the image can then be deconvolved by the usual methods.
These techniques are discussed and compared in greater detail in
\citet{stanimirovic2002}, beautifully applied to 21-cm mosaic maps of
the Small Molecular Cloud from Parkes and the ACTA. Two recently
developed, alternative approaches which fold single dish data into the
deconvolution process are the so-called TP2VIS
\citep{koda2011,koda2019} and SDINT \citep{rau2019} methods. One
attractive feature of these methods is that they naturally accommodate
deconvolution of the single-dish PSF (primary beam).

The relative weights of the single dish and interferometer data will
strongly affect the characteristics of the combined image. These can
be set or adjusted manually but the best quality images are obtained
when the relative weights are determined by the intrinsic noise
properties of each dataset.  Therefore it is important that the
sensitivities of the single dish and interferometric maps or cubes are
well matched.  Two useful criteria are to match the single dish and
interferometer sensitivities in overlapping regions of the aperture
plane \citep{kurono2009,koda2011,mason2014} and to obtain single dish
data which result in an overall distribution of $uv$-plane weights
which is smooth and approximately Gaussian \citep{rodriguez2008}. When
combining with data from modern interferometer arrays, the implied
requirement on the single dish map sensitivity is sufficiently
stringent that large aperture single dishes and/or focal plane arrays
are often advantageous. 

%An intuitive minimum criterion is to require the single dish map to
%provide the surface brightness sensitivity that the high-resolution
%map would provide on the spatial scale of the lower resolution beam if
%it had full spatial frequency sampling. This corresponds to
%\begin{equation}
%\sigma_{SD} \sim \sqrt{\frac{\Omega_{B,SD}}{\Omega_{B,int}}} \sigma_{int} 
%  = \frac{\theta_{B,SD}}{\theta_{B,int}} \sigma_{int} 
%\end{equation}
%where $\Omega_{B,SD}$ and $\Omega_{B,int}$ are the areas of the single
%dish and interferometric restoring beams, and the map sensitivities
%$\sigma_{SD}$ and $\sigma_{int}$ are in units of Janskies per
%restoring beam.  
%pseudo-vis \citep{koda2011}

An illustration of the improvement that can be made by adding
single-dish data to an interferometer map is shown in
Figure~\ref{galsimFeather}. This is a simulated
observation\footnote{This simulation and imaging exercise uses sample
  data and scripts available on {\tt http://casaguides.nrao.edu/}.}
of a nearby galaxy, $\sim 1'$ in size, using ALMA at $\lambda=0.9 \,
{\rm mm}$. ALMA consists of 50 12-m antennas operating as an
interferometer, with an additional 4 12-m antennas operated as single
dishes to provide total power; 12 7-m antennas operated as an
interferometer to bridge the gap in the aperture plane between them.
In this simulated observation, the ALMA 12-m primary beam is $19'' \,
({\rm FWHM})$; 67 pointings of the 12-m ALMA array on a hexagonal
lattice are used to cover the field of interest, and 23 pointings
of the 7-m array. These 12-m and 7-m interferometric data were
deconvolved using the wide-field CLEAN algorithm implemented in CASA
(\S~\ref{sec:widefield}). The resulting CLEAN image is shown in the
left panel. In spite of the considerable improvement in sensitivity to
large spatial scales that the 7-m array provides, substantial negative
bowls are still evident around the source.  The right panel shows the
image resulting from combining the interferometric and total power
maps using feather.

\articlefigure{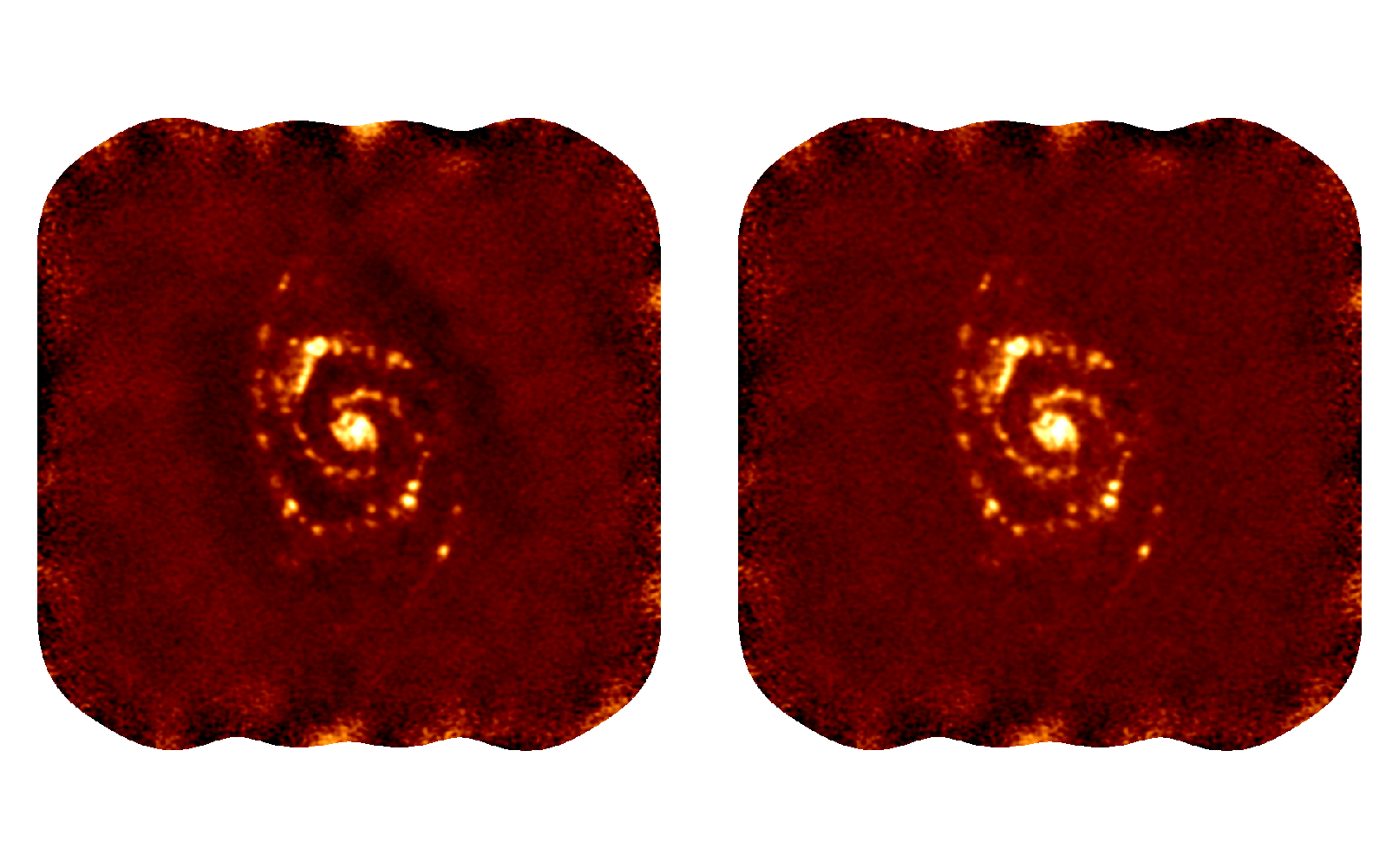}{galsimFeather}{\emph{Left}:
  Simulated ALMA 12m+7m $\lambda=0.9\, {\rm mm}$ interferometric
  mosaic of a galaxy $1'$ in size (ALMA 12-m primary beam: $19''$
  FWHM). \emph{Right}: after adding 12m total power data using the
  feather algorithm.}

\section{Practical Issues}
\label{sec:practical}

Following is a brief summary of some of the practical effects that
need to be considered when planning and analyzing mosaicked
interferometric observations and single dish observations intended to
support them.
\begin{itemize}

\item Choose an appropriate mosaic sampling strategy
  (\S~\ref{sec:sampling}). For high-fidelity imaging of extended
  emission, a hexagonal mosaic with pointing centers spaced by
  $\lambda/\sqrt{3} D$ is preferable.  Faster sky coverage can be
  achieved with sparser coverage, although this will not perform as
  well for retrieving short spacings, and will require more accurate
  knowledge of the primary beam. 

\item Many effects can cause the interferometer imaging
  characteristics to vary from one mosiac pointing to another.  This
  will tend to degrade the resulting image quality, depending on the
  reconstruction approach adopted.  For instance, the $uv$ coverage
  will change due to the Earth's rotation; flagging; the system
  temperature--- hence noise--- can vary; at short wavelengths the
  tropospheric phase can vary; and at long wavelengths ionospheric
  effects will vary.  Multiple coverages of the mosaic will tend to
  average such variations out, improving image quality, at the price
  of lower observing efficiency.

\item Minimizing antenna pointing errors is more important for
  interferometric mosaics than for single pointings. In the
  single-pointing case, the strongest emission is typically in the
  center of the beam, which is relatively flat; in the case of a
  mosaic, there is generally emission over the entire field, including
  areas where the primary beam has a steep gradient.

\item If single dish data is required, it is desirable that the dish
  diameter be at least $1.5 \times$ the minimum baseline in the
  interferometer, and preferably more. This is required to provide
  adequate $uv$ coverage and to be able to accurately link the
  calibration of the two instruments.

\item Single dish pointing errors can introduce spurious high spatial
  frequency structure to the map. These can often be alleviated by
  lightly smoothing.

\item The single dish antenna may have a non-trivial beam
  (\emph{e.g.}, due to an error beam, shadowing of the primary,
  \emph{etc.}) requiring deconvolution.

\item Depending on the calibration and imaging algorithms used the
  single dish map may have accurate information over only a limited
  range of spatial frequencies, not all the way down to $(u,v)=0$. This
  is particularly true for continuum data, which by definition lack a
  spectral dimension to help distinguish systematic effects from
  astronomical signal.

\item Ensure that the single dish map is sufficiently sensitive to
  provide useful information, and has a guard band of at least a few
  single-dish beams around the source.

\end{itemize}

% nyquist V

% things that cause varying noise and uvcoverage: ionosphere, atmosphere, flagging

\acknowledgements{The National Radio Astronomy Observatory is a
  facility of the National Science Foundation operated by Associated
  Universities, Inc.  This text is based on lectures given by the
  author at the 14th, 15th, and 16th NRAO Synthesis Imaging Summer
  Schools, and I thank the organizing committees for doing the hard
  work of getting together and running these workshops. I am also
  grateful to those who covered this topic at previous summer schools,
  especially Jurgen Ott and Steve Myers.  I thank Urvashi Rau and
  Crystal Brogan for helpful discussions, and Jeff Mangum and Adele
  Plunkett for comments on this manuscript.}

\bibliography{mosaicTutorial}

%Ekers \& Rots 1979
%Peterson Mathes 1965
%the white book
%the other white book

\end{document}